\documentclass{raa}

\usepackage{mathptmx}

\usepackage{ae,aecompl}

\usepackage{graphicx}	
\usepackage{amsmath}	
\usepackage{amssymb}	
\usepackage{pifont}
\usepackage{stmaryrd}
\usepackage{txfonts}
\usepackage{url}
\usepackage{subfigure}
\usepackage{longtable}
\usepackage{lscape}
\usepackage{multirow}
\usepackage{color}
\usepackage{textcomp}

\usepackage{natbib}
\usepackage[colorlinks,linkcolor=red,anchorcolor=green,citecolor=blue]{hyperref}

\newcommand\adeg{\mbox{$^\circ$}}%

\newcommand{\HII}{H {\small{II}} }
\newcommand{\kms}{{\rm km~s}^{-1}}
\newcommand{\Msun} {M_{\sun}}

\newcommand{\mjyb}{{\rm mJy~beam}^{-1}}
\newcommand{\Mjysr}{{\rm MJy~sr}^{-1}}

\newcommand{\jyb}{{\rm Jy~beam}^{-1}}

\begin{document}

\title{Searching for initial stage of massive star formation around the H II region G18.2-0.3}


   \volnopage{Vol.0 (201x) No.0, 000--000}      
   \setcounter{page}{1}          

   \author{Chuan-Peng Zhang
      \thanks{cpzhang@nao.cas.cn}
   \and Jing-Hua Yuan
   \and Jin-Long Xu
   \and Xiao-Lan Liu
   \and Nai-Ping Yu
   \and Nan Li
   \and Li-Ping He
   \and Guo-Yin Zhang
   \and Jun-Jie Wang
   }


   \institute{National Astronomical Observatories, Chinese Academy of Sciences, 100012 Beijing, PR China}

   \date{Received~~2016 month day; accepted~~2017~~month day}

\abstract{Sometimes the early star formation can be found in cold and dense molecular clouds, such as infrared dark cloud (IRDC). Considering star formation often occurs in clustered condition, \HII regions may be triggering a new generation of star formation, so we can search for initial stage of massive star formation around \HII regions. Based on that above, this work is to introduce one method of how to search for initial stage of massive star formation around \HII regions. Towards one sample of the \HII region G18.2-0.3, multiwavelength observations are carried out to investigate its physical condition. In contrast and analysis, we find three potential initial stages of massive star formation, suggesting that it is feasible to search for initial stage of massive star formation around \HII regions.
 }

   \authorrunning{Chuan-Peng Zhang et al.}            
   \titlerunning{Searching for initial stage of massive star formation}  

   \maketitle


\section{Introduction}    
\label{sect:intro}

High-mass star formation is playing an important role in forming the Milky Way \citep{whit1994,full2005}. Regions of massive star formation are on average more distant than the sites of low-mass star formation. The early stage of clustered star formation is characterized by dense, parsec-scale filamentary structures interspersed with complexes of dense cores (< 0.1 pc cores clustered in complexes separated by $\sim$ 1 pc) with masses from about 10 to 100 $\Msun$ \citep{Battersby2014}. So far, we still have poor knowledge about the process of early high-mass star formation, due to the initial stage of high-mass star formation is one of the most difficult detections and studies by our instruments \citep{Motte2007,Pillai2007,Pillai2011}. Another reason is that we have no enough sample in prestellar stage. Therefore, to get more samples we propose one method of searching for initial stage of massive star formation around \HII regions.

IRDCs are often suggested as the precursors to massive stars and stellar clusters \citep{Rathborne2007,Rathborne2010,Sanhueza2012,Jimenez2014,Wang2014}. Lots of studies \citep[e.g.,][]{Rathborne2007,Rathborne2010,Zhang2017} about IRDCs have been focusing on the earliest massive star formation. However, other condition can also breed young stellar objects (YSOs), such as surrounding \HII region and supernova remnant (SNR). \HII regions are manifestations of newly formed massive stars that are still embedded in their natal molecular clouds \citep{wals1997,poma2009,Zhang2014}. Dust in the molecular cloud renders \HII regions observable only at radio, infrared, and sub-millimeter wavelengths \citep{Churchwell2002}. The central star of an \HII region is believed to have ceased accreting matter and to have settled down for a short lifetime on the main sequence \citep{Hofner2002}. In addition, \HII regions are almost always accompanied by molecular clouds on their borders. The Orion Nebula, for example, is merely a conspicuous ionized region on the nearby face of a much larger dark cloud; the \HII region is almost entirely produced by the ionization provided by a single hot star \citep{wals1997,Churchwell2002,poma2009}. The studies of infrared dust bubbles associated with \HII regions have revealed some triggered star formation in the ringlike shell \citep[e.g.,][]{s51,n68,n131,Zhang2016,Yuan2014}.

The goal of this work is to introduce one method of searching for initial stage of massive star formation around the \HII region G18.2-0.3, which has an area of $0.4\adeg\times0.4\adeg$ centered at $l=18.2\adeg,\, b=-0.3\adeg$ (see Figs. \ref{Fig:870um} and \ref{Fig:nvss-8um}). The \HII region G18.2-0.3 consists of the SNR G18.1-0.1 \citep{Green2009}, infrared dust bubbles N21 and N22 \citep{Churchwell2006}, and the \HII regions G018.149-00.283 \citep{Kolpak2003}, G18.197-00.181 \citep{Lockman1989}, and G18.237-0.240 \citep{Paron2013}. The distance is around 4 kpc \citep{Paron2013}. This target is selected based on that  massive stars are usually born in clusters probably from material of the same molecular cloud, which then produce, along their evolution, neighbouring \HII regions, interstellar bubbles and SNRs that can interact with the parental cloud \citep{Paron2013}.

In this work, a multiwavelength observations are carried out to investigate the physical condition of the \HII region G18.2-0.3. The molecular line $^{13}$CO (1-0) and dust continuum including 8.0 $\mu$m, 70 $\mu$m, 870 $\mu$m, 21 cm, and \textit{Herschel} data are adopted to study the \HII region. This paper is arranged as follows. Section \ref{sect:archive} presents the data used in this work. Section \ref{sect:results} shows the results of the data analysis. In Section \ref{sect:discu}, we discuss how to search for initial stage of massive star formation around a well-selected \HII region complex, the associations of the initial stage of massive star formation with \HII regions nearby, and the property of associated massive star formation. Finally, a summary is presented in Section \ref{sect:summary}.

\section{Archival data}
\label{sect:archive}

\subsection{Dust continuum}

The combined dust continuum data comprise \textit{Spitzer} IRAC 8.0 $\mu$m \citep[1$\sigma = 78~\mu$Jy;][]{benj2003,chur2009}. The resolution at 8.0 $\mu$m is $\sim$2.0$''$. The InfraRed Array Camera (IRAC) is one of three focal plane instruments on the \textit{Spitzer}\footnote{This work is based partly on observations made with the \textit{Spitzer} Space Telescope, which is operated by the Jet Propulsion Laboratory, California Institute of Technology under a contract with NASA.} Space Telescope. IRAC is a four-channel camera that provides simultaneous $5.2' \times 5.2'$ images at 3.6, 4.5, 5.8, and 8.0 $\mu$m. The Multiband Imaging Photometer for \textit{Spitzer} (MIPS) produced imaging and photometry in three broad spectral bands, centered nominally at 24, 70, and 160 $\mu$m, and low-resolution spectroscopy between 55 and 95 $\mu$m. The resolution at 24 $\mu$m is $\sim$6$''$. The \textit{Herschel}\footnote{\textit{Herschel} is an ESA space observatory with science instruments provided by European-led Principal Investigator consortia and with important participation from NASA.} Space Observatory is a 3.5 meter telescope observing the Far-Infrared and Submillimeter Universe. The imaging bands for the Photo detector Array Camera and Spectrometer (PACS) were centered at 70, 100, and 160 $\mu$m \citep[1$\sigma_{\rm 70~{\mu}m}$ = 20 $\Mjysr$, 1$\sigma_{\rm 160~{\mu}m}$ = 20 $\Mjysr$;][]{Poglitsch2010,Molinari2016}. The resolution at 70 and 160 $\mu$m is about 8.4$''$ and 13.5$''$, respectively. SPIRE 250, 350, and 500 $\mu$m \citep[1$\sigma_{\rm 250~{\mu}m}$ = 10 $\Mjysr$, 1$\sigma_{\rm 350~{\mu}m}$ = 4 $\Mjysr$, and 1$\sigma_{\rm 500~{\mu}m}$ = 2 $\Mjysr$;][]{Griffin2010,Molinari2016} have a spatial resolution of about 18.1$''$, 24.9$''$, and 36.4$''$, respectively. ATLASGAL\footnote{The ATLASGAL project is a collaboration between the Max-Planck-Gesellschaft, the European Southern Observatory (ESO) and the Universidad de Chile.} 870 $\mu$m \citep[1$\sigma = 54\,\mjyb$;][]{Schuller2009,Csengeri2014} is the APEX Telescope Large Area Survey of the Galaxy, an observing programme with the LABOCA bolometer array at APEX, located at 5100 m altitude on Chajnantor, Chile. Its spatial resolution at 870 $\mu$m is about 19$''$. The radio continuum data at 21 cm, with a synthesized beam of about 45$''$, was extracted from observations for the 1.4 GHz NRAO VLA Sky Survey \citep[NVSS;][]{cond1998}.

\begin{figure*}
\centering
\includegraphics[width=0.60\textwidth, angle=0]{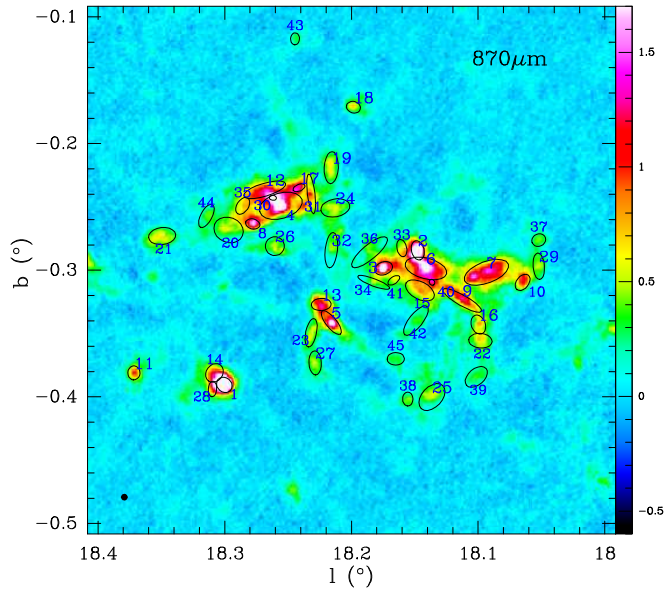}
\caption{870 $\mu$m dust continuum emission overlaid by the extracted Gaussian clumps for the \HII region G18.2-0.3. The unit of 870 $\mu$m color bar is in $\jyb$. The beam size is indicated at the bottom-left corner.}
\label{Fig:870um}
\end{figure*}

\subsection{Molecular line}

The molecular line data is accessed from the Milky Way Galactic Ring Survey (GRS), which was performed by a Boston University (BU) and Five College Radio Astronomy Observatory (FCRAO) collaboration \citep{Jack2006}. Using the SEQUOIA multi-pixel array receiver on the FCRAO 14 m telescope, $^{13}$CO (1-0) survey of the inner Galaxy was conducted. The GRS offers a sensitivity of < 0.4 K, spectral resolution of 0.2 $\kms$, and angular resolution of 46$''$ with sampling 22$''$. The original intensities are on antenna temperature scale $T_{\rm A}^{*}$. To convert this to main beam temperatures $T_{\rm MB}$, $T_{\rm A}^{*}$ is divided by the main beam efficiency of 0.48. The GILDAS\footnote{http://www.iram.fr/IRAMFR/GILDAS/} software package was used to reduce the molecular line data.

\begin{table*}
\caption{Parameters of extracted clumps}
\label{tab_obs} \centering 
\begin{tabular}{rrrrrrrrrrrr} 
\hline \hline
No.  &  offset-x & offset-y  &  Distance  &  $I_{870\mu m}$  &  FWHMx  &  FWHMy  &  $R_{\rm eff}$  &  $T_{\rm dust}$  &  $N_{\rm H_2}$  &  Mass  &  Lum.  \\  
     &  $''$  &  $''$  &  kpc  &  $\jyb$  &  $''$  &  $''$  &  pc  &  K  &  $\rm cm^{-2}$  &  $\Msun$  &  $L_\odot$  \\  
\hline
1	&	361.00	&	-325.78	&	$	3.22	\pm	0.20	$	&	4.06	&	46.47	&	40.71	&	0.34	&	30.4	&	2.54E+22	&	717	&	39185	\\
2	&	-188.47	&	56.70	&	$	3.52	\pm	0.23	$	&	3.08	&	56.13	&	35.96	&	0.38	&	30.0	&	1.99E+22	&	442	&	46717	\\
\textbf{3}	&	\textbf{-91.60}	&	\textbf{5.75}	&	$	\textbf{3.40}	\pm	\textbf{0.19}	$	&	\textbf{2.23}	&	\textbf{51.63}	&	\textbf{36.80}	&	\textbf{0.36}	&	\textbf{23.1}	&	\textbf{1.90E+22}	&	\textbf{405}	&	\textbf{16746}	\\
4	&	200.44	&	183.52	&	$	4.92	\pm	0.28	$	&	2.22	&	95.34	&	62.25	&	0.92	&	24.7	&	1.76E+22	&	1411	&	34990	\\
\textbf{5}	&	\textbf{57.24}	&	\textbf{-148.65}	&	$	\textbf{3.40}	\pm	\textbf{0.19}	$	&	\textbf{1.80}	&	\textbf{83.31}	&	\textbf{30.06}	&	\textbf{0.41}	&	\textbf{14.7}	&	\textbf{8.27E+22}	&	\textbf{1243}	&	\textbf{5256}	\\
6	&	-211.70	&	5.40	&	$	3.52	\pm	0.23	$	&	1.74	&	91.04	&	59.00	&	0.63	&	31.7	&	1.49E+22	&	1108	&	100109	\\
7	&	-383.42	&	-5.90	&	$	3.45	\pm	0.20	$	&	1.46	&	98.13	&	60.93	&	0.65	&	21.8	&	2.18E+22	&	1325	&	28660	\\
\textbf{8}	&	\textbf{280.40}	&	\textbf{131.60}	&	$	\textbf{4.92}	\pm	\textbf{0.28}	$	&	\textbf{1.43}	&	\textbf{41.16}	&	\textbf{28.44}	&	\textbf{0.41}	&	\textbf{22.2}	&	\textbf{1.23E+22}	&	\textbf{219}	&	\textbf{3705}	\\
9	&	-314.69	&	-85.81	&	$	3.52	\pm	0.23	$	&	1.21	&	55.52	&	29.84	&	0.35	&	24.2	&	1.17E+22	&	294	&	7963	\\
10	&	-486.34	&	-34.27	&	$	3.40	\pm	0.19	$	&	1.11	&	52.32	&	34.65	&	0.35	&	21.3	&	1.30E+22	&	295	&	4045	\\
11	&	617.90	&	-291.80	&	$	3.36	\pm	0.18	$	&	1.03	&	39.90	&	34.39	&	0.30	&	23.6	&	1.06E+22	&	252	&	3702	\\
12	&	246.07	&	228.85	&	$	3.45	\pm	0.20	$	&	0.98	&	84.07	&	36.55	&	0.46	&	21.2	&	1.77E+22	&	802	&	9845	\\
13	&	85.82	&	-97.22	&	$	3.40	\pm	0.19	$	&	0.90	&	56.42	&	34.91	&	0.37	&	20.9	&	1.86E+22	&	561	&	7053	\\
14	&	389.10	&	-291.83	&	$	3.22	\pm	0.20	$	&	0.85	&	56.43	&	46.55	&	0.40	&	26.1	&	2.46E+22	&	870	&	31342	\\
15	&	-194.52	&	-57.25	&	$	3.45	\pm	0.20	$	&	0.80	&	88.83	&	49.44	&	0.55	&	26.2	&	1.07E+22	&	772	&	38883	\\
16	&	-360.47	&	-154.53	&	$	3.52	\pm	0.23	$	&	0.74	&	56.03	&	39.06	&	0.40	&	20.4	&	1.25E+22	&	467	&	4788	\\
17	&	148.76	&	234.61	&	$	3.45	\pm	0.20	$	&	0.73	&	35.63	&	18.85	&	0.22	&	25.9	&	1.27E+22	&	152	&	4829	\\
18	&	-5.73	&	463.46	&	$	3.45	\pm	0.20	$	&	0.70	&	40.44	&	32.84	&	0.30	&	26.3	&	6.31E+21	&	134	&	5164	\\
19	&	57.23	&	291.80	&	$	3.52	\pm	0.23	$	&	0.62	&	91.29	&	39.82	&	0.51	&	28.4	&	6.27E+21	&	331	&	16238	\\
20	&	349.03	&	114.47	&	$	4.92	\pm	0.28	$	&	0.60	&	85.59	&	70.99	&	0.93	&	24.1	&	5.58E+21	&	741	&	14289	\\
21	&	537.83	&	97.23	&	$	4.92	\pm	0.28	$	&	0.59	&	77.39	&	47.99	&	0.73	&	17.1	&	2.54E+22	&	687	&	5414	\\
22	&	-366.17	&	-200.28	&	$	3.52	\pm	0.23	$	&	0.58	&	66.77	&	41.72	&	0.45	&	18.4	&	1.75E+22	&	543	&	4130	\\
23	&	114.41	&	-177.38	&	$	3.40	\pm	0.19	$	&	0.56	&	82.53	&	30.62	&	0.41	&	22.2	&	7.50E+21	&	416	&	5116	\\
24	&	45.78	&	177.34	&	$	3.45	\pm	0.20	$	&	0.56	&	83.22	&	50.81	&	0.54	&	27.0	&	6.50E+21	&	431	&	19329	\\
25	&	-228.84	&	-360.44	&	$	3.40	\pm	0.19	$	&	0.54	&	89.77	&	57.12	&	0.59	&	22.6	&	6.47E+21	&	603	&	9579	\\
26	&	217.41	&	68.65	&	$	4.92	\pm	0.28	$	&	0.54	&	53.61	&	53.53	&	0.64	&	29.6	&	3.39E+21	&	208	&	11795	\\
27	&	102.99	&	-263.20	&	$	3.40	\pm	0.19	$	&	0.54	&	68.33	&	35.35	&	0.41	&	24.1	&	7.15E+21	&	235	&	6126	\\
28	&	394.80	&	-337.58	&	$	3.22	\pm	0.20	$	&	0.52	&	44.27	&	24.17	&	0.26	&	27.8	&	1.02E+22	&	296	&	16110	\\
29	&	-532.10	&	11.45	&	$	3.40	\pm	0.19	$	&	0.52	&	75.50	&	32.48	&	0.41	&	17.8	&	1.54E+22	&	412	&	3056	\\
30	&	223.14	&	205.98	&	$	4.92	\pm	0.28	$	&	0.51	&	25.54	&	20.91	&	0.28	&	23.4	&	2.10E+22	&	197	&	3311	\\
31	&	114.43	&	217.43	&	$	3.45	\pm	0.20	$	&	0.51	&	53.25	&	20.92	&	0.28	&	27.3	&	8.67E+21	&	162	&	7306	\\
32	&	57.19	&	57.23	&	$	3.45	\pm	0.20	$	&	0.48	&	80.62	&	35.17	&	0.45	&	26.1	&	5.30E+21	&	291	&	11778	\\
33	&	-143.03	&	62.94	&	$	3.52	\pm	0.23	$	&	0.47	&	50.19	&	27.60	&	0.32	&	31.4	&	6.90E+21	&	194	&	20932	\\
34	&	-62.94	&	-34.34	&	$	3.40	\pm	0.19	$	&	0.47	&	96.35	&	20.90	&	0.37	&	24.3	&	7.65E+21	&	243	&	6882	\\
35	&	308.97	&	183.08	&	$	4.92	\pm	0.28	$	&	0.44	&	53.17	&	32.45	&	0.50	&	21.2	&	1.14E+22	&	447	&	3848	\\
36	&	-51.49	&	51.50	&	$	3.45	\pm	0.20	$	&	0.43	&	88.75	&	41.03	&	0.50	&	27.6	&	5.45E+21	&	381	&	19025	\\
37	&	-532.11	&	85.83	&	$	3.40	\pm	0.19	$	&	0.42	&	39.80	&	35.38	&	0.31	&	17.7	&	1.45E+22	&	222	&	1603	\\
38	&	-160.20	&	-366.19	&	$	3.40	\pm	0.19	$	&	0.42	&	38.88	&	29.76	&	0.28	&	25.9	&	5.90E+21	&	118	&	3917	\\
39	&	-354.74	&	-303.24	&	$	3.40	\pm	0.19	$	&	0.41	&	73.45	&	45.71	&	0.48	&	21.2	&	8.40E+21	&	443	&	5025	\\
40	&	-228.86	&	-34.33	&	$	3.45	\pm	0.20	$	&	0.41	&	23.88	&	19.84	&	0.18	&	28.9	&	1.16E+22	&	85	&	5164	\\
41	&	-120.16	&	-28.61	&	$	3.45	\pm	0.20	$	&	0.39	&	38.45	&	19.49	&	0.23	&	27.2	&	6.98E+21	&	110	&	6328	\\
42	&	-183.10	&	-143.05	&	$	3.40	\pm	0.19	$	&	0.40	&	82.49	&	38.12	&	0.46	&	21.3	&	8.26E+21	&	411	&	7237	\\
43	&	160.20	&	657.98	&	$	3.40	\pm	0.19	$	&	0.38	&	33.81	&	25.32	&	0.24	&	23.1	&	6.82E+21	&	104	&	1807	\\
44	&	411.95	&	154.48	&	$	4.92	\pm	0.28	$	&	0.38	&	72.15	&	32.29	&	0.58	&	21.6	&	7.15E+21	&	260	&	3931	\\
45	&	-125.87	&	-251.76	&	$	3.40	\pm	0.19	$	&	0.38	&	49.04	&	35.85	&	0.35	&	25.9	&	5.89E+21	&	171	&	6064	\\
\hline
\end{tabular}
\begin{itemize}
 \item
The offset (0, 0) is located at the position of $l=18.2\adeg$, $b=0.3\adeg$.
 \item 
The clumps 3, 5, and 8 are potentially three initial stages of massive star formation.
\end{itemize}
\end{table*}

\section{Analysis}
\label{sect:results}

\subsection{Clump extraction}
\label{sect_extraction}

A typical terminology \citep[e.g.,][]{Bergin2007} for clump has a physical size of 0.3 -- 3 pc with a mass of about 50 -- 500 $\Msun$. Therefore, based on the derived effective radius in Table \ref{tab_obs}, the extracted objects are called as clump in this work.

\begin{figure*}
\centering
\includegraphics[width=0.60\textwidth, angle=0]{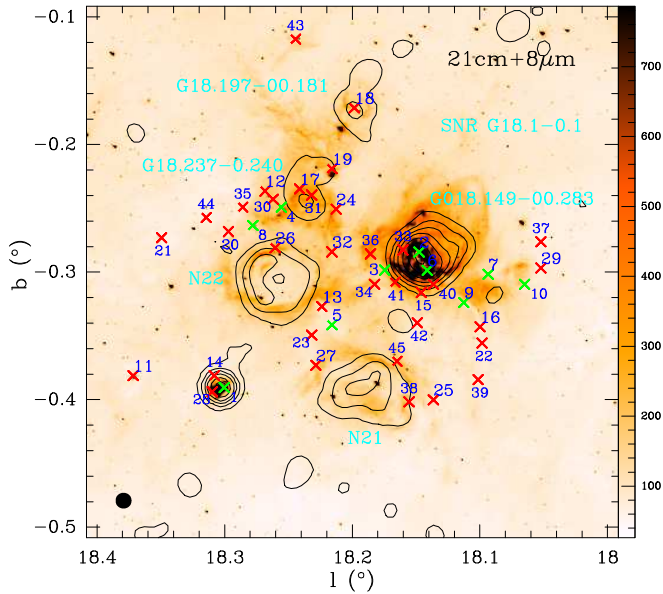}
\caption{21 cm dust continuum contours superimposed on image of 8 $\mu$m emission for the \HII region G18.2-0.3. Several \HII regions are indicated with their names. The contour levels of 21 cm emission are, respectively, 3$\sigma$, 20$\sigma$, 50$\sigma$, 100$\sigma$, 200$\sigma$, 350$\sigma$, and 520$\sigma$ with $\sigma$ = 0.002 $\jyb$. The unit of 8 $\mu$m color bar is in $\Mjysr$. The crosses with numbers indicate the peak positions of extracted massive clumps (see Fig. \ref{Fig:870um}). Particularly, the green crosses show top 10 most massive clumps.}
\label{Fig:nvss-8um}
\end{figure*}

The potential massive clumps are extracted with \textit{Gaussclumps} procedure \citep{Stutzki1990,Kramer1998} in GILDAS software package in 870 $\mu$m map, assuming that the flux density of each clump is in Gaussian distribution. \textit{Gaussclumps} can be used to fit a 2-dimensional clump locally to the maximum of the input cube. It then subtracts this clump from the cube, creating a residual map, and then continues with the maximum of this residual map. The procedure is repeated until a stop criterion is met, for instance when the maximum of the residual maps drops below the 3 sigma level. We just consider the clumps with peak intensity of 870 $\mu$m emission above 6$\sigma$. The measured parameters are also listed in Tables \ref{tab_obs}, and indicated with clump size in Fig. \ref{Fig:870um} and also with crosses in Fig. \ref{Fig:nvss-8um}. The measured FWHMx and FWHMy have been convolved with beam size.

\begin{figure*}
\centering
\includegraphics[width=0.60\textwidth, angle=0]{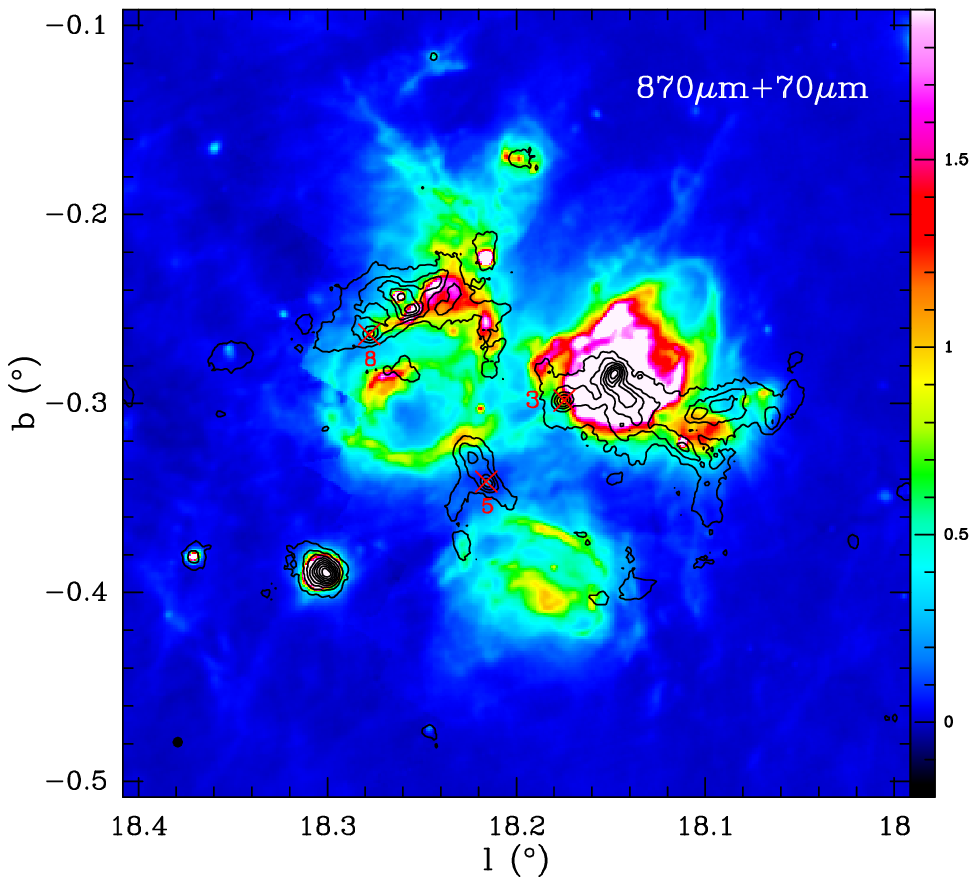}
\caption{870 $\mu$m dust continuum contours superimposed on image of 70 $\mu$m emission for the \HII region G18.2-0.3. The contour levels of 870 $\mu$m emission start at 6$\sigma$ in steps of 8$\sigma$ with $\sigma$ = 0.054 $\jyb$, and its beam size is indicated at the bottom-left corner. The red crosses with numbers show three potential initial stages of massive star formation. The unit of 70 $\mu$m color bar is in MJy sr$^{-1}$.}
\label{Fig:870-70um}
\end{figure*}

\begin{figure*}
\centering
\includegraphics[width=0.60\textwidth, angle=0]{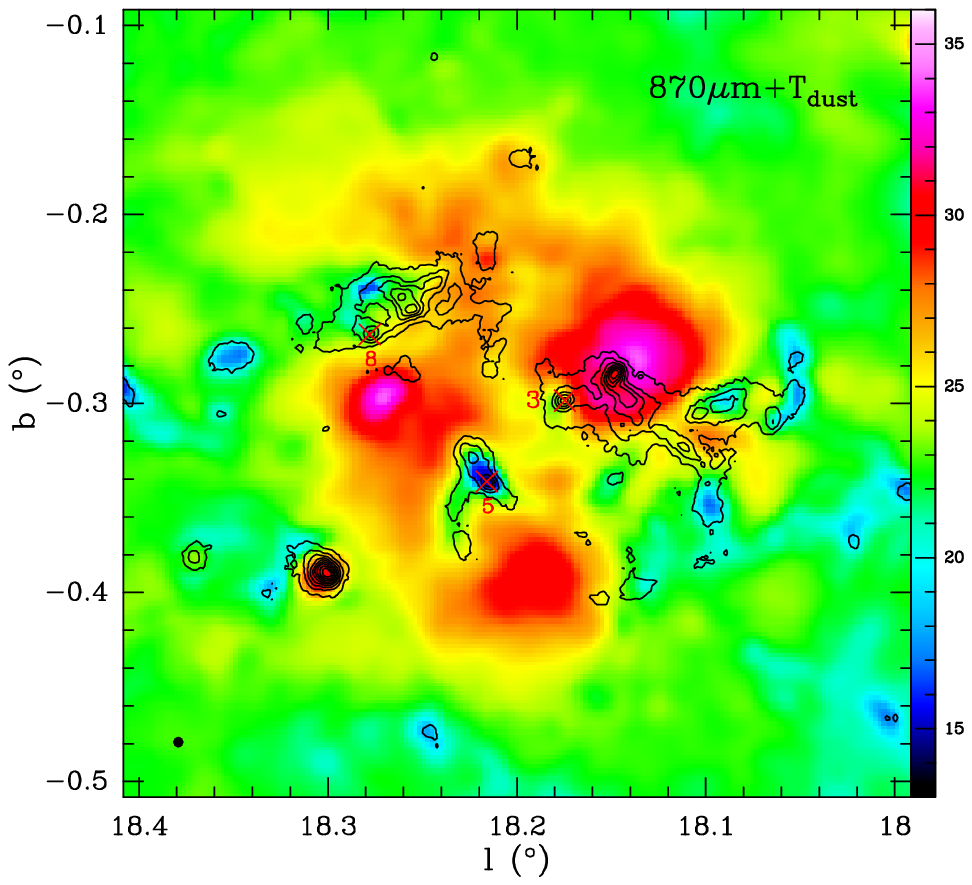}
\caption{870 $\mu$m dust continuum contours superimposed on image of dust temperature $T_{\rm dust}$ for the \HII region G18.2-0.3. The contour levels of 870 $\mu$m emission start at 6$\sigma$ in steps of 8$\sigma$ with $\sigma$ = 0.054 $\jyb$, and its beam size is indicated at the bottom-left corner. The red crosses with numbers show three potential initial stages of massive star formation. The unit of $T_{\rm dust}$ color bar is in K.}
\label{Fig:Tdust}
\end{figure*}

\subsection{Initial stage of massive star formation}

Prestellar core is often suggested as the precursor of massive star formation. Prestellar cores represent a somewhat denser and more centrally-concentrated population of cores which are starless but self-gravitating \citep{Andre2009}. They are typically detected in sub-millimeter dust continuum emission and dense molecular gas tracers, often seen in absorption at mid- to far-infrared wavelengths \citep{Tan2013,Chitsazzadeh2014,Wang2011,Wang2014,Cyganowski2014,Kong2016}. However, prestellar cores are really difficult to detect, due to that they almost have nothing infrared emission.

In this work, we define infrared quiet clump as the initial stage of massive star formation. Infrared quiet clumps have all physical properties, such as cold ($T < 18$ K), dense ($n > 10^5$ cm$^{-3}$), just excluded the low luminosity \citep[$S_{\rm 24 \, \mu m}$ < $(\frac{\rm 1.7\,kpc}{D})^2 \times 15.0$ Jy;][]{Motte2007,Andre2009,Russeil2010}. As a potentially early stage of molecular clouds, infrared dark clouds have been discovered two decades ago as dark patches in mid-infrared (MIR) images of the Galactic plane \citep{Perault1996,Egan1998} and many studies of the physical conditions within them have been conducted recently \citep{Pillai2006,Pillai2007,Wyrowski2008,Pillai2012,Zhang2017}. \HII regions do not represent the earliest stage of massive star formation, as is often claimed \citep[e.g.,][]{Churchwell2002}. Star formation in the Milky Way always takes place in clusters and groups within large molecular clouds \citep{Aikawa2005,Hennebelle2008,Andre2009,Pagani2013}. It is likely that the triggered early star formation by \HII region nearby has different views from the IRDCs, such as the triggering or breeding condition. Therefore we will search for initial stage of massive star formation around an \HII region complex, as is the rationale of this work.

Three clumps are proposed as potentially initial stage of massive star formation. They are, respectively, clumps No. 3, 5, and 8 listed in Table \ref{tab_obs}, and presented in Figure \ref{Fig:870-70um}. We further searched for compact point sources in MIPS 24 $\mu$m \citep{care2009,Gutermuth2015}, PACS 70 $\mu$m \citep{Molinari2016}, and ATLASGAL 870 $\mu$m \citep{Csengeri2014} catalogs, and found that there exist counterparts for clumps No. 3, 5, and 8. The counterparts at 24, 70, 870 $\mu$m are listed in Table \ref{tab_point}. Based on the definition of infrared quiet clumps ($S_{\rm 24 \, \mu m}$ < $(\frac{\rm 1.7\,kpc}{D})^2 \times 15.0$ Jy), the infrared property of clumps No. 3, 5, and 8 meet this criteria. 

\begin{table*}
\caption{Point source information at 24, 70, and 870 $\mu$m}
\label{tab_point} \centering 
\footnotesize 
\begin{tabular}{ccccccccc} 
\hline \hline
No.  &$l$&$b$& MIPS$_{\rm 24\,{\mu}m}$ & $S_{\rm 24\,{\mu}m}$  & PACS$_{\rm 70\,{\mu}m}$ &  $S_{\rm 70\,{\mu}m}$ & ATLASGAL$_{\rm 870\,{\mu}m}$ & $S_{\rm 870\,{\mu}m}$ \\  
     &$\adeg$ & $\adeg$ & & mJy &  &  Jy & & Jy    \\  
\hline
3 & 18.178179 & -0.299098 & MG018.1751-00.2985 & 214.881 & HIGALPB018.1782-0.2991 & 6.569 & AGAL018.174-00.299  & 45.38 \\
5 & 18.215827 & -0.341628 & MG018.2157-00.3417 & 773.266 & HIGALPB018.2158-0.3416 & 3.993 & AGAL018.214-00.342  & 9.47 \\
8 & 18.276361 & -0.263550 & MG018.2758-00.2636 & 463.319 & HIGALPB018.2764-0.2635 & 7.604 & AGAL018.278-00.262  & 22.86 \\
\hline
\end{tabular}
\end{table*}

\subsection{Dust temperature and column density}
\label{sect:temperature}

The high-quality \textit{Herschel} data cover a large wavelength range from 70 to 500 $\mu$m making it practical to obtain dust temperature maps of the \HII region via fitting the SED to the multi-wavelength images on a pixel-by-pixel basis. Firstly we have followed \citet{Wang2015} to perform Fourier-Transfer (FT) based on background removal. In this method, the original images have firstly transformed into Fourier domain and separated into the low and high spatial frequency components, and then inversely Fourier transfered back into image domain. The low-frequency component corresponds to large-scale background/foreground emission, while the high-frequency component reserves the emission of interest. Detailed descriptions of the FT-based background removal method can be found in \citet{Wang2015}. After removing the background/foreground emission, we have re-gridded the pixels onto the same scale of 11.5$''$, and convolved the images to a circular Gaussian beam with $\mathrm{FWHM = 36.4''}$ which corresponds to the measured beam of \textit{Herschel} observations at 500 $\mu$m \citep{Traficante2011}. The intensities at multi-wavelengths of each pixel have been modeled as
	\begin{eqnarray}
	\begin{aligned}
	\label{equa_Su}
S_\nu=B_\nu(T) (1-e^{-\tau_\nu})
	\end{aligned}
	\end{eqnarray}
where the Planck function $B_\nu(T)$ is modified by optical depth
	\begin{eqnarray}
	\begin{aligned}
	\label{equa_tau}
\tau_\nu = \mu_\mathrm{H_2}m_\mathrm{H}\kappa_\nu N_\mathrm{H_2}/R_\mathrm{gd}.
	\end{aligned}
	\end{eqnarray}
Here, $\mu_\mathrm{H_2}=2.8$ is the mean molecular weight adopted from \citet{Kauffmann2008}, $m_\mathrm{H}$ is the mass of a hydrogen atom, $N_\mathrm{H_2}$ is the column density, $R_\mathrm{gd}=100$ is the gas to dust ratio. The dust opacity $\kappa_\nu$ can be expressed as a power law of frequency with
	\begin{eqnarray}
	\begin{aligned}
	\label{equa_kappa}
\kappa_\nu=5.0\left(\frac{\nu}{600~\mathrm{GHz}}\right)^\beta~\mathrm{cm^2g^{-1}},
	\end{aligned}
	\end{eqnarray}
where $\kappa_\nu(\mathrm{600~GHz})=5.0~\mathrm{cm^2g^{-1}}$ adopted from \citet{Ossenkopf1994}. The dust emissivity index has been fixed to be $\beta=1.75$ according to \citet{Battersby2011}. The free parameters are the dust temperature $T_{\rm dust}$ and column density $N_\mathrm{H_2}$.

The final resulted dust temperature map, which has a spatial resolution of 36.4$''$ with a pixel size of 11.5$''$, is shown in Fig. \ref{Fig:Tdust}. Other parameters are listed in Table \ref{tab_obs}. We have to admit that the derived dust temperatures are over-estimated due to contamination from the emission of \HII regions nearby.

\subsection{Luminosity}

The total energy radiated from an object per second (named as Boltzmann luminosity) can be expressed by 
	\begin{eqnarray}
	\begin{aligned}
	\label{equa_luminosity}
L_{\rm clump} = 4 \pi d^2 \Omega_{\rm pix} \sum I_{\rm int},
	\end{aligned}
	\end{eqnarray}
where $d$, $\Omega$, and $I_{\rm int}$ are the distance, solid angle, and flow of energy out of a surface at each source, respectively. The $I_{\rm int}$ for each pixel can be estimated using the resultant dust temperature and column density (see Section \ref{sect:temperature}). The luminosities of the sources with distance measurements were calculated by integrating the frequency-integrated intensities between 10$^2$ and 10$^5$ GHz within the Gaussian ellipses. The derived luminosities are listed in Table \ref{tab_obs}. We have to admit that the derived Boltzmann luminosity are also over-estimated like dust temperature due to contamination from the emission of \HII regions nearby.

\begin{figure*}
\centering
\includegraphics[width=0.60\textwidth, angle=0]{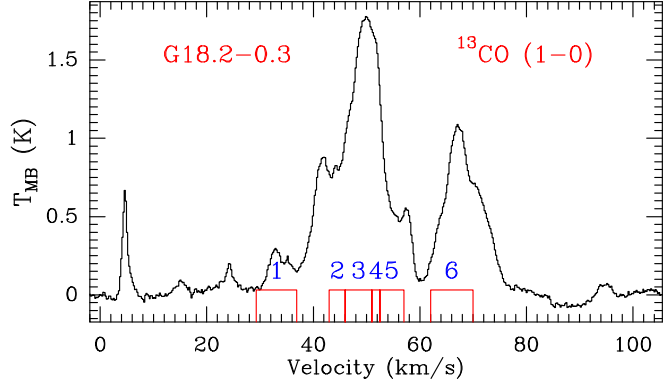}
\caption{Averaged $^{13}$CO line within the whole \HII region G18.2-0.3. The red windows with numbers indicate six different velocity components (see Table \ref{tab_velocity} and Fig. \ref{Fig:chan_map}) to be further investigated.}
\label{Fig:spectrum}
\end{figure*}

\begin{figure*}
\centering
\includegraphics[width=0.90\textwidth, angle=0]{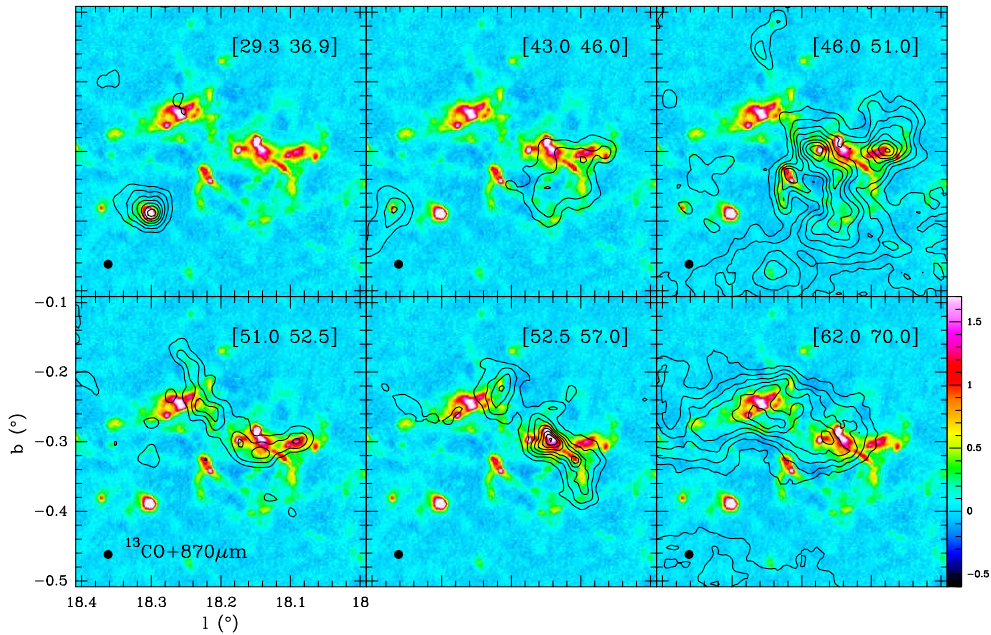}
\caption{Integrated-velocity channel maps of $^{13}$CO lines superimposed on 870 $\mu$m emission for the \HII region G18.2-0.3. The integrated velocity ranges are indicated in each panel with unit of $\kms$. The contour levels of the $^{13}$CO lines start at 3$\sigma$ in steps of 1$\sigma$ with $\sigma$ = 2.08 ${\rm K}\, \kms$. The beam size $^{13}$CO data is indicated at the bottom-left corner. The unit of 870 $\mu$m color bar is in $\jyb$.}
\label{Fig:chan_map}
\end{figure*}

\begin{figure*}
\centering
\includegraphics[width=0.70\textwidth, angle=0]{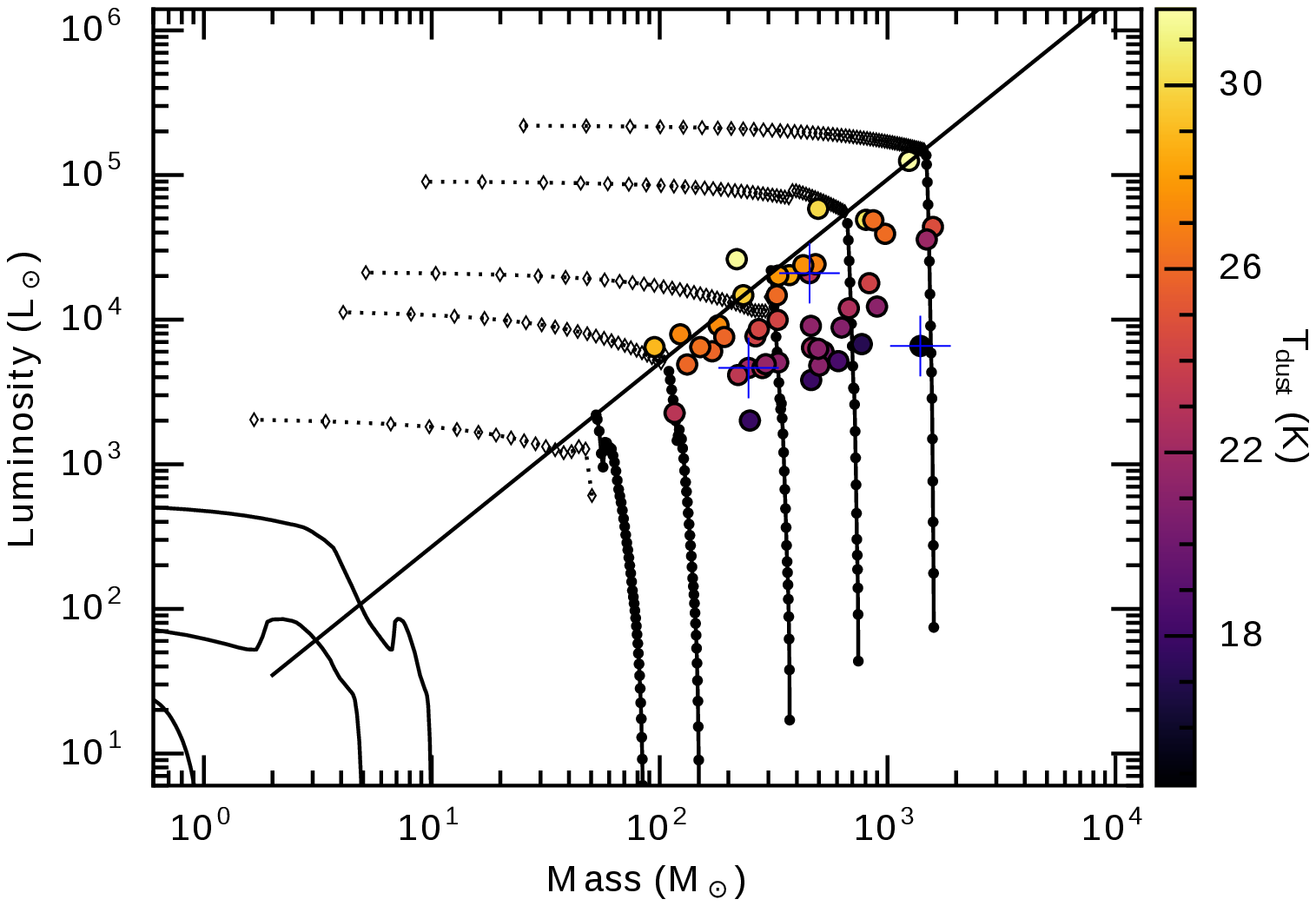}
\caption{Luminosity-Mass diagram for sources presented in Table \ref{tab_obs}. Lines and tracks are from \citet{Saraceno1996} \& \citet{Molinari2008}, and depict the situation for the low-mass/low-luminosity (below the line) and high-mass/high-luminosity (above the line) regimes. The filled circles with different color indicate the sources with different dust temperatures. The three blue crosses indicate the found potential initial stages (with No. 3, 5, and 8 in Table \ref{tab_obs}) of massive star formation. }
\label{Fig:relationML}
\end{figure*}

\begin{figure*}
\centering
\includegraphics[width=0.60\textwidth, angle=0]{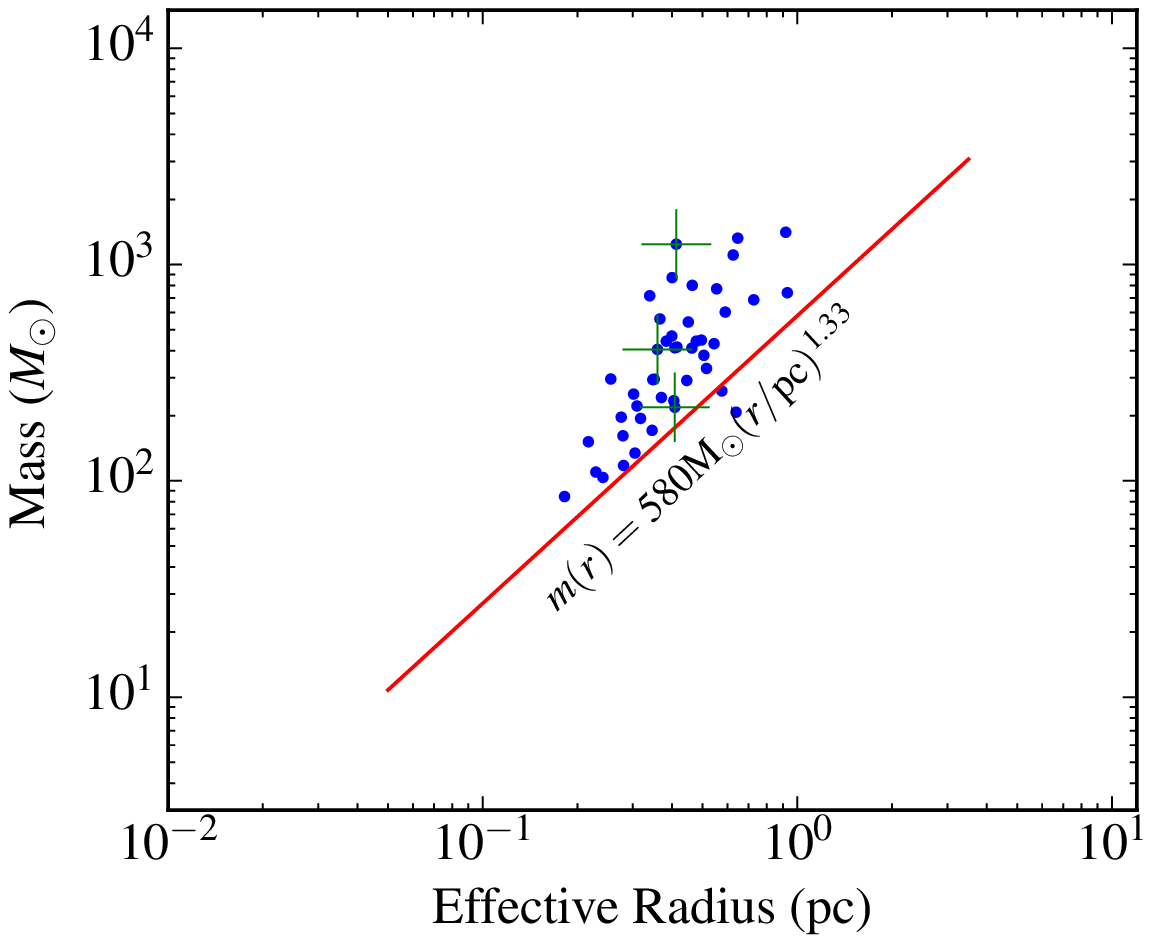}
\caption{$Mass$-$radius$ distributions of Gaussian clumps extracted from $Gaussclumps$. The masses and effective radius are listed in Table \ref{tab_obs}. The red line delineates the threshold introduced by \citet{Kauffmann2010} separating the regimes under which high-mass stars can form (above) or not (below line). The clump masses in blue points (see Table \ref{tab_obs}) are derived from multiwavelength \textit{Herschel} data. The three green crosses indicate the found potential initial stages (with No. 3, 5, and 8 in Table \ref{tab_obs}) of massive star formation.}
\label{Fig:mass_size}
\end{figure*}

\subsection{Clump mass and virial mass}

We just use \textit{Herschel} data along with the derived dust temperatures in Section \ref{sect:temperature} to estimate masses of these extracted clumps. The mass is given by the integral of the column densities across the source,
	\begin{eqnarray}
	\begin{aligned}
	\label{equa_kappa}
M=\mu_{\rm H_2}m_{\rm H}d^2\int N_{\rm H_2} d\Omega,
	\end{aligned}
	\end{eqnarray}
where $d$ and $\Omega$ are the distance and solid angle of the source, respectively. These corresponding and derived parameters are listed in Table \ref{tab_obs}.

The virial theorem can be used to test whether one clump is in a stable state. Under the assumption of a simple spherical clump with a density distribution of $\rho$ = constant, if ignoring magnetic fields and bulk motions of the gas \citep{MacLaren1988,Evans1999},
	\begin{eqnarray}
	\begin{aligned}
	\label{equa_virial-mass}
    M_{\rm vir} \simeq 210\, r\, \Delta V^{2}\, (\Msun),
	\end{aligned}
	\end{eqnarray}
where $r$ is adopted with the clump effective radius in pc and $\Delta V$ (listed in Figure \ref{tab_clump}) is the full width at half-maximum line width in $\kms$. The $\Delta V$ was estimated with $^{13}$CO line. The spatial resolution of the $^{13}$CO data is partly larger than the sizes of individual clumps, so we just considered the $^{13}$CO spectrum within one pixel corresponding with the peaked position of each clump. The virial parameter $\alpha_{\rm vir}$ is defined by $\alpha_{\rm vir} = M_{\rm vir}/M$. For the found three initial stages of massive star formation, their virial masses and virial parameters are listed in Table \ref{tab_clump}. In such dense clumps, $^{13}$CO line becomes optically thick, so the virial mass is likely over-estimated.

\begin{table*}
\caption{The parameters of the found three massive clumps in initial stage}
\label{tab_clump} \centering 
\begin{tabular}{cccccc} 
\hline \hline
Clump  & $\Delta V$ &  $Mass$ & $M_{\rm vir}$  &  $\alpha_{\rm vir}$ & $t_{\rm clump}$   \\  
     &  $\kms$ & $\Msun$  &  $\Msun$  &  &  Myr     \\  
\hline
3  &  $6.09\pm0.06$ &  405  &  2804  &  6.92   & 1.7  \\
5  &  $3.90\pm0.08$ & 1243  &  1310  &  1.05   & 2.7  \\
8  &  $6.87\pm0.12$ &  219  &  4064  &  16.59  & 2.9  \\
\hline
\end{tabular}
\end{table*}

\subsection{Different velocity components}
\label{sect:velocity}

In Fig. \ref{Fig:spectrum}, we show the averaged $^{13}$CO line within the whole \HII region G18.2-0.3. The red windows with numbers indicate six different velocity components (see Table \ref{tab_velocity} and Fig. \ref{Fig:chan_map}) to be further investigated. The molecular clouds in different distances pile up into together, so that we often observe different velocity components in line of sight.

Dust continuum has no velocity information, so we can combine molecular line $^{13}$CO (1-0) to investigate the velocity correlation with the massive clumps. If the $^{13}$CO (1-0) emission in different integrated-velocity ranges has good correlation with any 870 $\mu$m dust continuum distribution, its corresponding velocity information can be obtained (see Figure \ref{Fig:chan_map}). 

Furthermore, the distances to the massive clumps was derived based on the Bayesian Distance Calculator\footnote{http://bessel.vlbi-astrometry.org/bayesian} \citep{Reid2016}, which leverages these results to significantly improve the accuracy and reliability of distance estimates to other sources that are known to follow spiral structure. \citet{Paron2013} listed distances of several \HII regions and one SNR, which are close to the derived distances with Bayesian Distance Calculator. We think that Bayesian Distance Calculator is more reliable, so it will be adopted in this work. Based on the analysis in Figure \ref{Fig:chan_map}, the distances of all velocity ranges for the \HII region G18.2-0.3 are listed in Table \ref{tab_velocity}.

In work of \citet{Paron2013}, however, the authors skipped the velocity component of [62.0 70.0] $\kms$, which is actually associated with the \HII region G18.237-0.240, as is unknown before. We have to note that the studied \HII region complex G18.2-0.3 is indeed consisted of many different and complicated components in line of sight, and the \HII region G18.237-0.240 (associated with the velocity component of [62.0 70.0] $\kms$) is located at a different arm from the other parts of the complex (see Table \ref{tab_velocity}).

\begin{table*}
\caption{Different velocity components and distances associated with G18.2-0.3}
\label{tab_velocity} \centering 
\begin{tabular}{l|cccccc} 
\hline \hline
Veolcity window   &  1 & 2 & 3 & 4 & 5 & 6   \\  
Velocity ($\kms$)  &  [29.3 36.9] & [43.0 46.0] & [46.0 51.0] & [51.0 52.5] & [52.5 57.0] & [62.0 70.0]   \\  
\hline
Line center ($\kms$) & 33.10      & 44.50 & 48.50 & 51.75 & 54.75 & 66.00 \\
Distance (kpc)     &  3.22(0.20)  & 3.36(0.18)  & 3.40(0.19)  & 3.45(0.20)  & 3.52(0.23)  & 4.92(0.28)    \\
Probability        &  0.95        & 1.00        & 1.00        & 0.91        & 0.78        & 0.64          \\
Spiral arm         &  ScN         &  ScN        &  ScN        &  ScN        &  ScN        & Nor           \\ 
\hline
\end{tabular}
\end{table*}

\section{Discussion}
\label{sect:discu}

\subsection{Evolutionary time in \HII regions and clump formation}

We searched for the NVSS catalog and obtained a total flux of $S_{\nu}$ = 4417.0, 710.9, 796.4 mJy at $\nu$ = 1.4 GHz for the \HII regions G018.148-00.283, N22, and N21, respectively \citep{cond1998}. The flux of stellar Lyman photons $N_{\rm LyC}$, absorbed by the gas in the \HII region, can be derived from the relation \citep{mezg1974} as
\begin{equation}
    \label{eq:nlcy}
   \left(\frac{N_{\rm LyC}}{\rm s^{-1}}\right) = \frac{4.761 \times 10^{48}}{a(\nu, T_{\rm e})} \left(\frac{\nu}{\rm
GHz}\right)^{0.1} \left(\frac{T_{\rm e}}{\rm K}\right)^{-0.45} \left(\frac{S_{\nu}}{\rm Jy}\right) \left(\frac{D}{\rm
kpc}\right)^2,
\end{equation}
where $a(\nu, T_{\rm e}) \sim 1$ is a slowly varying function tabulated by \citet{mezg1967}, the electron temperature of the \HII region is assumed to be $T_{e} \sim$ 8000 K, and $D$ is distance. The power exponent of $T_{e}$ is small, so the result does not depend strongly on the chosen $T_{e}$. Based on above, The derived Lyman-continuum ionizing photon flux and the equivalent star style \citep{pana1973} are listed in Table \ref{tab_hii}.

To check whether the evolutionary status of the \HII regions is old enough to trigger new generation star formation nearby, we can compare the evolutionary time scales between \HII regions and clump formation. We estimate their evolutionary status, using the model described by \citet{dyso1980} as
\begin{equation}
    \label{eq:hii}
t{_{\rm H\,II}}=\frac{4~R_s}{7~c_s}\left[\left(\frac{R}{R_s}\right)^{7/4}-1\right],
\end{equation}
where $R$ is the radius of \HII regions obtained from the NVSS catalog (see Table \ref{tab_hii}), $c_s$ = 10 $\kms$ is the sound velocity in the ionized gas  and $R_s$ is the radius of the Str\"omgren sphere given by $R_s=(3N_{\rm Lyc}/4\pi n_0^2\alpha_B)^{1/3}$, where $N_{\rm Lyc}$ is the number of ionizing photons emitted by the star per second, $n_0$ = $\sim(1.0\pm0.5) \times 10^3$ cm$^{-3}$ is the original ambient density, and $\alpha_B$ = 2.6 $\times 10^{-13}$ cm$^3$ s$^{-1}$ is the hydrogen recombination coefficient to all levels above the ground level. Finally, we derive the dynamical age of each \HII region, which is also listed in Table \ref{tab_hii}.

We estimate the fragmentation time of the three early high-mass candidates (clumps 3, 5, 8) potentially triggered by an \HII region nearby according to the theoretical model from \citet{whit1994}:
\begin{equation}
    \label{eq:clump}
    t_{\rm clump}=1.56 \left(\frac{\alpha_s}{0.2}\right)^{7/11}
    \left(\frac{N_{\rm Lyc}}{10^{49}}\right)^{-1/11}
    \left(\frac{n_0}{10^3}\right)^{-5/11}~\left[\rm Myr\right].
\end{equation}
The turbulent velocity $\alpha_s$ can be estimated with line width in Table \ref{tab_clump}. Finally the derived fragmentation times for the three clumps are listed in Table \ref{tab_clump}.

Based on derived results in Tables \ref{tab_clump} and \ref{tab_hii}, we find that the evolutionary time of the \HII region is longer than the fragmentation time of the three clumps 3, 5, 8. The fragmentation time is inferred by considering the uncertainty in the total Lyman continuum photon flux and turbulent velocity. Hence, the evolutionary status of the \HII regions seem to be responsible for the star formation activities around the \HII regions.

\subsection{The associations of initial stage of massive star formation with the \HII region nearby}

 Are the massive clump formations triggered by the \HII region nearby in Figure \ref{Fig:nvss-8um}? It has been proposed that the formation of \HII regions can trigger a new generation of star formation \citep[e.g.,][]{poma2009,wats2010}. In triggered star formation, one of several events might occur to compress a molecular cloud and initiate its gravitational collapse. Molecular clouds may collide with each other, or a nearby supernova explosion can be a trigger, sending shocked matter into the cloud at very high speeds \citep{Prialnik2000}. The triggered star formation may also happen at the waist of bipolar \HII region \citep{Deharveng2015}, where the high density ionized material flows away from the central region and high density molecular material accumulates to form a torus of compressed material. \citet{Ojha2011} presented an embedded cluster along with three prominent clumps appearing to be sandwiched between the two evolved \HII regions S255 and S257, and suggested that the positions of the young sources inside the gas ridge at the interface of the two \HII regions favor a site of induced star formation.

 Carefully checking the positions (see Section \ref{sect:velocity}) of the three initial stages of massive star formation, we found that the clumps No. 3 and 8 are located at the border of the western \HII region. Maybe the clumps No. 3 and 8 were triggered to be formed by strong stellar winds from the \HII region nearby. The clump No. 5 is located at the intersection between two \HII regions of bubbles N21 and N22. The case of clump No. 5 is very similar to the sandwiched star formation. It is probably that the clump No. 5 was born from the compression of the \HII regions of bubbles N21 and N22. Therefore, the \HII region nearby may be triggering a new generation of star formations, which will be studied in detail in our follow-up works.
 
\begin{table*}
\caption{The parameters of the \HII regions}
\label{tab_hii} \centering 
\begin{tabular}{lcccccc} 
\hline \hline
\HII  &  $R$  & $D$ &  $S_{\rm 1.4 GHz}$ & log($N_{\rm LyC}$) & Stage   &  $t_{\rm H\,II}$    \\  
     & pc & kpc & mJy  & &  &  Myr   \\  
\hline
 G018.148-00.283 & 0.68   & $3.52\pm0.23$ &  4417.0  &  48.68 &  O7   & 2.7  \\
 bubble N22      & 0.78   & $4.92\pm0.28$ &   710.9  &  48.17 &  O8.5   & 4.7  \\
 bubble N21      & 0.75   & $3.40\pm0.19$ &   796.4  &  47.89 &  O9.5   & 5.2  \\
\hline
\end{tabular}
\end{table*}

\subsection{The property of associated massive star formation}

In Fig. \ref{Fig:mass_size}, we present the mass-size plane for the extracted clumps at 870 $\mu$m. Comparison with the high-mass star formation threshold of $m(r) > 580 {\Msun} (r/{\rm pc})^{1.33}$ empirically proposed by \citet{Kauffmann2010} allows us to determine whether these clumps are capable of giving birth to massive stars. The data points are distributed above the threshold (given by the red line in Fig. \ref{Fig:mass_size}) that discriminates between high and low mass star formation whose entries fall above and bellow the line, respectively, indicative of high-mass star-forming candidates. It appears that the most of clumps are high-mass star-forming candidates at 870 $\mu$m. Particularly, the potential three initial stages of massive star formation (marked with green crosses) are apparently located above the threshold, suggesting they are high-mass star candidates.

The derived virial masses $M_{\rm vir}$ and virial parameters $\alpha_{\rm vir}$ are listed in Table \ref{tab_clump} for the potential three initial stages of massive star formation. Of the three clumps, $\alpha_{\rm vir} > 1$ for clumps No. 3 and 8, suggesting that the clumps are not gravitationally bound, in a stable or expanding state, while $\alpha_{\rm vir} < 1$ for clump No. 5, suggesting that the clump is gravitationally bound, potentially unstable, and collapsing \citep{Hindson2013}. The interaction within each clump is deserved to further study, and to understand their initial stages in detail using higher spatial resolution instrument.

\subsection{How to search for initial stage of massive star formation around an \HII region?}

Since \HII regions may trigger a new generation of star formation \citep{chur2007,chur2008,wats2008,s51}, it is likely that one can obtain early star formation around \HII regions. Evidences have shown that the star formation in some different evolutionary stages can be found around \HII region, such as starless cores, hot cores, outflows, and protostars \citep{poma2009,zava2010,s51}.  Generally the \HII region has strong continuum emission at centimeter wavelength. We can use, e.g. 21 cm continuum, to trace an \HII region. High-mass star formation in early stage has a cold, dense, and dark condition. It is well-known that e.g. 870 $\mu$m continuum are proposed as one good tracer. Some \HII regions, such as hyper-compact \HII region and hot core, may be deeply embedded in cold and dense envelope \citep{Zhang2014}, however, they show very low luminosity. These objects do not belong to the initial stage of massive star formation. Therefore, some of these compact clumps at 870 $\mu$m are not the true earliest stage. We need to further remove these clumps with weak centimeter and infrared emissions, and to get relatively early stage of star formation.

In Figure \ref{Fig:nvss-8um}, the crosses with numbers are the extracted clumps associated with 870 $\mu$m and \HII regions. Checking their masses and luminosities in Figure \ref{Fig:relationML}, we found that they almost have relatively high masses and low luminosities. In addition, in Figure \ref{Fig:mass_size}, the mass-size relation shows that most of them are distributed above the high-mass threshold. Particularly in Figure \ref{Fig:870-70um} and Table \ref{tab_point}, three dense clumps No. 3, 5, and 8 at 870 $\mu$m have weak 24 and 70 $\mu$m emission. In other word, these tree dense clumps have very weak infrared emission, but with strong emission at 870 $\mu$m, so they can be suggested as infrared quiet clumps. Their dust temperatures for No. 3, 5, 8 are 23.1, 14.7, 22.2 K (Figure \ref{Fig:Tdust}), with masses of  405, 1243, 219 $\Msun$, respectively. In Figures \ref{Fig:relationML} and \ref{Fig:mass_size}, we have highlighted the three clumps with crosses and numbers. Their properties above suggests that they are potentially initial stages of massive star formation.

\section{Summary}
\label{sect:summary}

In previous works, the early star formations were often located within IRDCs. In this work, considering star formation is in clustered condition, \HII region may be triggering a new generation of star formation. It is likely that we can search for initial stage of massive star formation around \HII regions. Therefore, this work is to present a method of how to search for initial stage of massive star formation around an \HII region. 

Towards the \HII region G18.2-0.3, we carry out a multiwavelength observations to investigate its dust temperature, luminosity, mass, density, the related velocity components, and evolutionary time. By contrast and analysis, finally we find three (in 45 clump candidates associated with the \HII region G18.2-0.3) potential initial stages of massive star formation, suggesting that it is feasible to search for initial stage of massive star formation around \HII regions.

\section*{Acknowledgements}
\addcontentsline{toc}{section}{Acknowledgements}

We wish to thank the anonymous referee for comments and suggestions that improved the clarity of the paper. C.-P. Zhang is supported by the Young Researcher Grant of National Astronomical Observatories, Chinese Academy of Sciences. This work is partly supported by the National Key Basic Research Program of China (973 Program) 2015CB857100, and National Natural Science Foundation of China 11503035, 11363004, 11403042. This publication makes use of molecular line data from the Boston University-FCRAO Galactic Ring Survey (GRS). The GRS is a joint project of Boston University and Five College Radio Astronomy Observatory, funded by the National Science Foundation under grants AST-9800334, AST-0098562, \& AST-0100793.

\bibliographystyle{raa}
\bibliography{references}

\end{document}